\documentstyle[prb,aps,twocolumn]{revtex}
\input epsf
\begin{document}
\draft

\title{Adiabatic steering and determination of dephasing rates  in  double dot qubits}

\author{T. Brandes $^1$, F. Renzoni $^2$, and R. H. Blick $^3$}

\address{ $^1$ Department of Physics, University of Manchester Institute of
Science and Technology (UMIST), P.O. Box 88, Manchester M60 1QD, United 
Kingdom} 

\address{$^2$ Laboratoire Kastler-Brossel, Departement de Physique de l'Ecole
Normale Sup\'erieure, 24 rue Lhomond, 75231, Paris Cedex 05, France}

\address{$^3$ Center for NanoScience and Sektion Physik, 
Ludwig--Maximiliam--Universit\"at, Geschwister--Scholl--Platz 1, 
80539 M\"unchen, Germany}
\date{\today{ }}
\maketitle
\begin{abstract}
We propose a scheme to prepare arbitrary superpositions of quantum states
in  double quantum--dots irradiated by coherent microwave pulses. 
Solving the equations of motion for the dot density matrix, we find 
that dephasing rates for such superpositions can be quantitatively infered from
additional electron current pulses that appear
due to a controllable breakdown of coherent population trapping in the dots.
\end{abstract}
%\pacs{73.23.H, 73.50.Pz, 73.40.Gk}
\pacs{
73.21.La,    %Quantum dots  
32.80.Qk,    %Coherent control of atomic interactions with photons 
85.35.Gv     %Single electron devices  
}

\section{Introduction}
% CHANGED COMPLETELY
One key point for the implementation  of quantum logic gates in quantum dots 
is the preparation of arbitrary superpositions $|\Psi_f\rangle$ of two (or more)
electron eigenstates. Such superpositions would constitute a qubit basis in an
artificial semiconductor structure that can be coupled to the external world by
leads and therefore be accessed by {transport spectroscopy}. Transport experiments
have already successfully revealed a number of quantum coherent effects due to static
potential coupling \cite{Vaartetal95,Blietal98b,FujetalTaretal} or coupling to
microwave radiation \cite{Oosetal98,Blietal98a,Holetal00} in double quantum dots, 
and their use as qubits  that are controlled by gate--voltages or magnetic fields
has been suggested \cite{BL00} recently. Unfortunately, % to the best of our
at present there are no data at all for dephasing rates $\gamma$ of 
superpositions $|\Psi_f\rangle$ in such systems.
The knowledge of $\gamma$, in particular for lowest eigenstate superpositions that
are expected to be most stable, is crucial to determine the feasibility of quantum
dot based qubits and to choose the time scale for logic operations needed for, e.g.,
quantum computation \cite{Bouwmeester}. 

In this article, we suggest a scheme to {\em determine} $\gamma$ from time--dependent 
transport measurements through coupled quantum dots. We show that it is
indeed possible to prepare an arbitrary coherent superposition of ground states using
stimulated Raman adiabatic passage \cite{bergmann} with two microwave pulses irradiated
on a double dot in the Coulomb blockade regime. 
Solving the equations of motion for the dot density matrix, it turns out that
two subsequent pairs of microwave pulses filter out the degree of decoherence
in the form of a detectable electron current peak, the strength of which directly
depends on $\gamma$. 

% CHANGED
Apart from conventional photon--assisted tunneling experiments with single microwave sources,
%\cite{Kouetal94,Oosetal98,Blietal98a,Holetal00}, 
recent {\em two}--source microwave 
techniques have turned out to be an extremely versatile tool
to investigate both ground and excited states in single dots \cite{Qinetal01}. 
Furthermore, different groups \cite{Bonetal98,ALB00,Hohetal00} have already identified
{\em three--level systems} 
%with two ground and one excited state
as potential candidates to establish and control
low--frequency coherence in solid state structures.
%*** In this work, we examine $\Lambda$--type couplings in tunnel coupled dots, examples of which
%are shown in Fig. \ref{dotscheme}, together with the double--dot scheme where
%two ground states are coupled to a common excited state by two microwave fields. ***

%%%%%%%%%%%%%%%%%%%%%%%%%%%%%%%%%%%%%%%%%%%%%%%%%%%%%%%%%%%%%%%%%%
\begin{figure}[ht]
\setlength{\unitlength}{1in}
%\begin{center}
\begin{picture}(4,1.5)
%\put(1.8,1.6){\epsfxsize 0.9in \epsfbox{ddot_1.ps}}
\put(-0.1,0.4){\epsfxsize 1.1 in \epsfbox{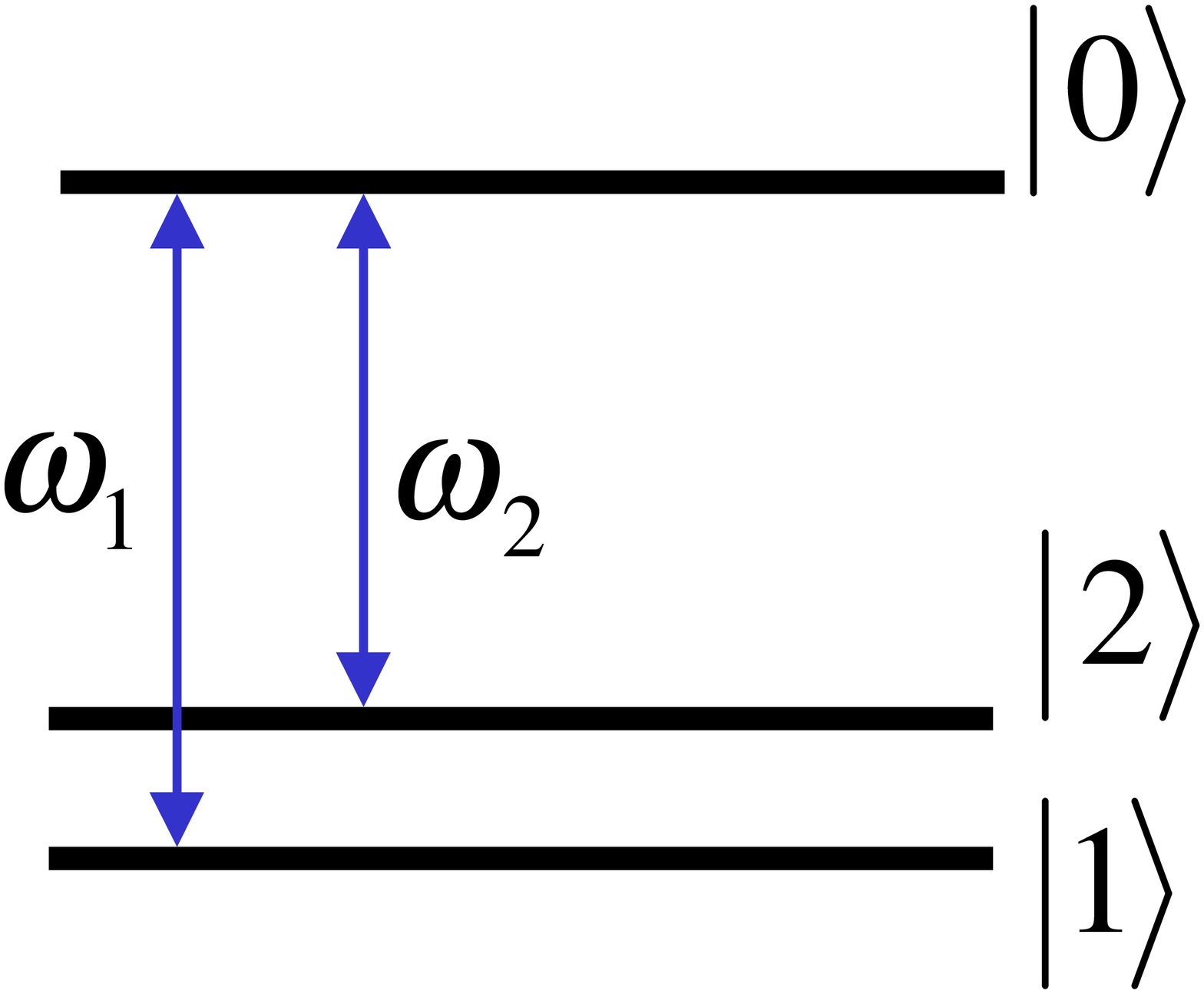}}
\put(1.0,0.4){\epsfxsize 1.1 in \epsfbox{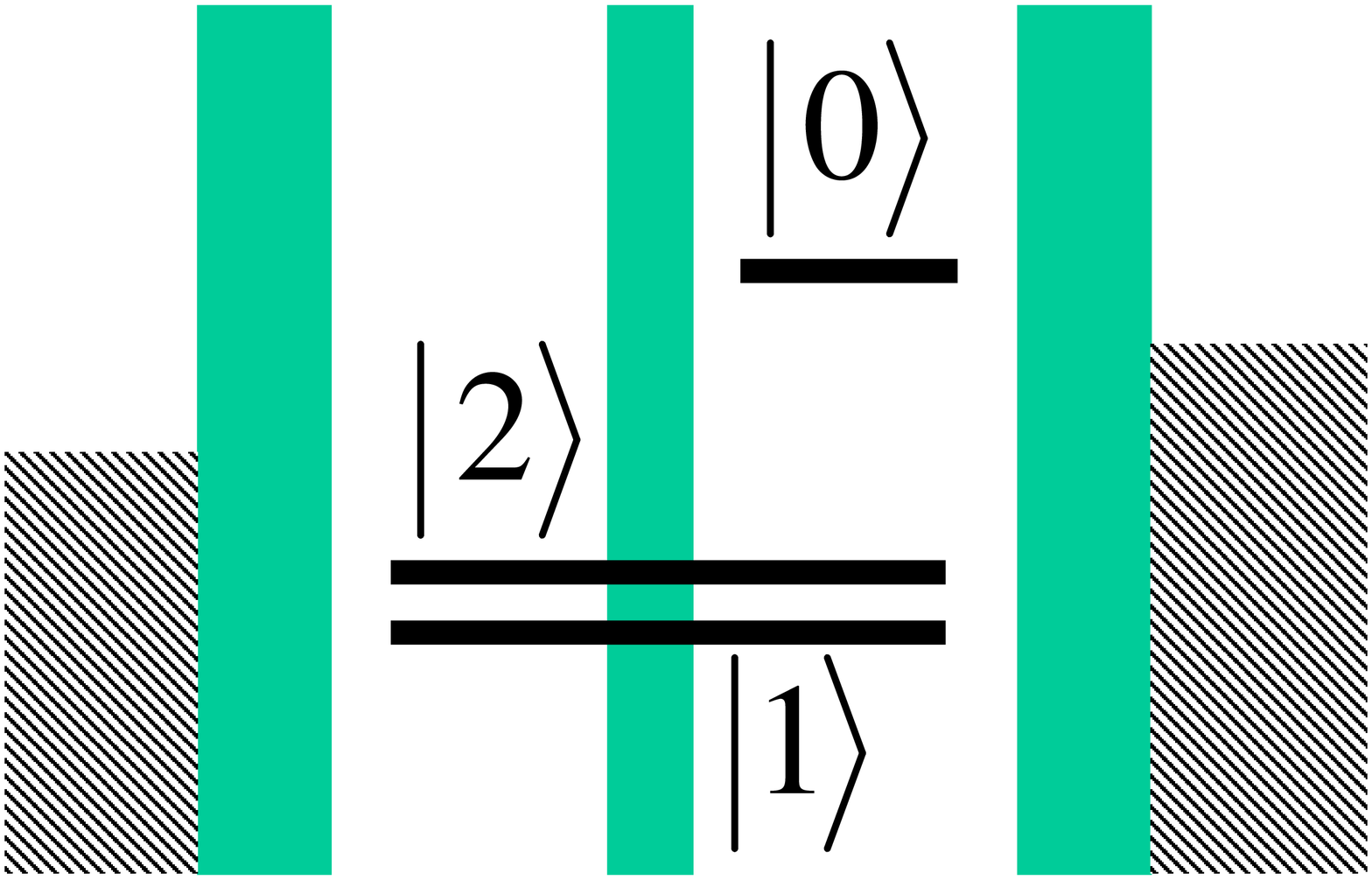}}
\put(2.2,0.4){\epsfxsize 0.8 in \epsfbox{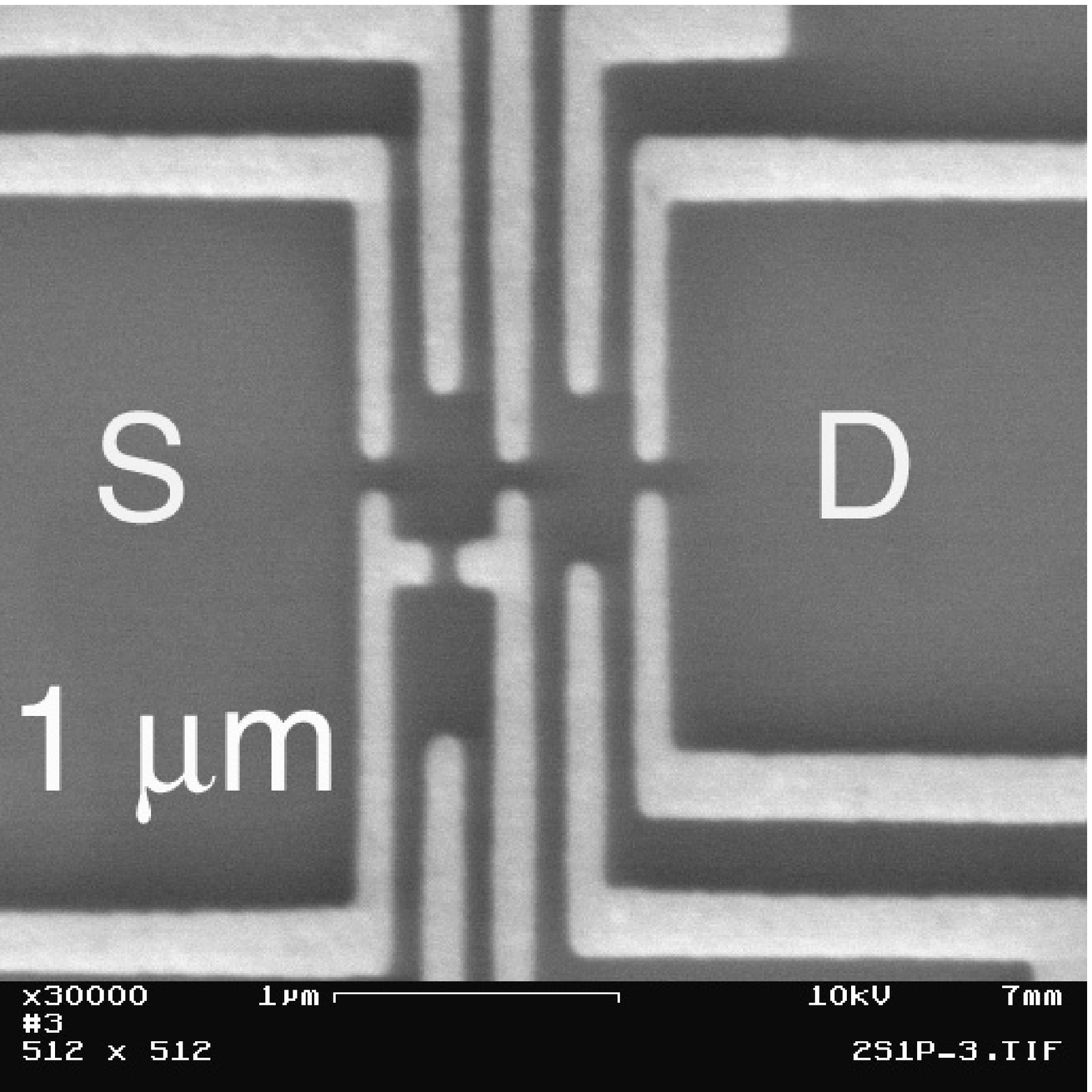}}
%\mbox{\epsfxsize 3in \epsfbox{fig1.ps}}
%\mbox{\epsfxsize 1in \epsfbox{ddotscheme.ps}}
%\put(-1.2,1.7){\epsfxsize 0.8in \epsfbox{ddotscheme.ps}}
%\end{center}
\end{picture}
\caption{$\Lambda$--configuration (left), three--level scheme in 
coupled two--dot system (center), and lateral three--dot device in a 
GaAs/AlGaAs heterostructure. The $1 \mu$m mark gives a size scale for the structure.}
\label{dotscheme}
\end{figure}
%%%%%%%%%%%%%%%%%%%%%%%%%%%%%%%%%%%%%%%%%%%%%%%%%%%%%%%%%%%%%%%%%%

\section{Adiabatic Transfer Scheme}
Our scheme to prepare and analyze superpositions is based on previously discussed
dark resonance states in lateral double dots \cite{BR00} that can be probed by linear
and non--linear transport spectroscopy \cite{Blietal98b,FujetalTaretal,Blietal98a}:
A left and a right dot are coupled to external leads with chemical potentials such that
electrons can tunnel out to the right only via an excited state $|0\rangle$ in the 
right dot and tunnel in  from the left and the right only via the two double dot ground 
states $|1\rangle$ and $|2\rangle$, see Fig. \ref{dotscheme}.
The latter are the symmetric  and antisymmetric superpositions of the left and the right 
tunnel--coupled dot groundstates for one additional electron.
In the regime of weak coupling, this corresponds to `ionic' binding, while strong coupling
can be termed `covalent binding'.
Microwave--induced `dark' stable superpositions 
$|\psi_f\rangle$ of $|1\rangle$ and $|2\rangle$ then constitute the qubit and 
can be formed by adiabatic transfer as discussed in the following.
In this work, we concentrate on $\Lambda$--type couplings in coupled $N$--dot devices for $N=2$ 
(double dots, Fig. \ref{dotscheme} left and center). For 
%gate voltage controled 
multiple dot devices with $N>2$ (Fig. \ref{dotscheme} right), though more difficult to 
control, we expect even more 
possibilities to manipulate energy levels and the shape of wave functions.

We assume the dot initially prepared in $|1\rangle$, which can be easily
achieved emptying level $|2\rangle$ by driving the $|2\rangle\leftrightarrow
|0\rangle$ transition with a microwave field. Once preparation
of the dot in $|1\rangle$ is achieved, the dot 
is set to a ground state superposition of $|1\rangle$ and $|2\rangle$
by irradiation with two electric fields of the form 
\begin{eqnarray}
{\bf E}_i(t)=\vec{{\cal E}}_i(t)\cos (\omega_i t+\varphi_i), \quad i=1,2,   
\end{eqnarray}
with microwave frequencies 
$\omega_i$ of the order of the transition frequencies 
$\delta\varepsilon_i/\hbar$ and slowly varying pulse--shaped 
amplitudes $\vec{{\cal E}}_i(t)$. 
%({\em Comment on phase difference see below})
The latter give rise to time--dependent  
matrix elements $\Omega_P(t)\propto |\vec{{\cal E}}_1(t)|$ and $\Omega_S(t) \propto |\vec{{\cal E}}_2(t)|$
for the transitions $0\leftrightarrow 1$ (P) and
$0\leftrightarrow 2$ (S) that induce the adiabatic transfer of the 
qubit from the initial state to the desired superposition.
Without loss of generality, we assume real Rabi frequencies  \cite{vita} 
\begin{mathletters}
\begin{eqnarray}
\Omega_P(t)&=&\Omega^0 \sin \theta e^{-(t-\tau)^2/T^2},\\
\Omega_S(t)&=&\Omega^0 \left( e^{-t^2/T^2} + \cos \theta 
e^{-(t-\tau)^2/T^2}\right).
\end{eqnarray}
\label{eq:pulse}
\end{mathletters}
%for the preparation of ground state superpositions with real coefficients. 
Here, $\tau$ and $T$ are the pulse delay and pulse duration, respectively.
The precise Gaussian form of Eq.~(\ref{eq:pulse}) is not a strict 
requirement for the process of adiabatic transfer and has been chosen 
only for convenience. 

The Stokes microwave pulse $S$ couples
$|2\rangle$ to $|0\rangle$, before a second pulse (the pump pulse $P$),
partially overlapping with $S$, couples $|1\rangle$ to $|0\rangle$ \cite{marte,vita}.
If the pulses terminate simultaneously with a constant ratio of their
amplitudes, the dot is left in a superposition 
\begin{equation}\label{psif}
|\psi_f\rangle=\cos\theta |1\rangle - \sin\theta |2\rangle,
\end{equation}
where the mixing angle $\theta$ is determined by the ratio with which the pump
and Stokes pulses terminate.  The process is robust against experimental details 
such as the delay between pulses, or pulse areas. The only strict requirement is 
two--photon Raman resonance $\delta_R\equiv\delta_2-\delta_1$ where
$\delta_j = \omega_j-\delta\epsilon_j/\hbar$ is the one photon detuning.
The phase difference $\varphi_1-\varphi_2$ of the two electric fields is either
$0$ or $\pi$ so that by changing  $|\vec{{\cal E}}_1|$ and $|\vec{{\cal E}}_2|$ one covers the whole range 
of $\theta$ values.

\section{Model}
\subsection{Interaction Hamiltonian and Density Matrix Equations}
We now investigate the preparation of superpositions $|\psi_f\rangle$ and their
stability with respect to dephasing numerically. In the dipole and rotating wave
approximation, 
%CHANGED
as appropriate for near-resonant excitation,
%CHANGED
the time--dependent interaction Hamiltonian is
% CHANGED into array
\begin{eqnarray}
V_{\rm AL}(t) &=& -{\hbar\over 2}\left[
\Omega_P(t) e^{-i\omega_Pt} |0\rangle\langle 1| +
\Omega_S(t) e^{-i\omega_St} |0\rangle\langle 2|\right] \nonumber\\
&+& h.c.~.
\end{eqnarray}
%
% CHANGED 
%

For simplicity, we assume identical tunneling rates $\Gamma$ for tunneling
of electrons into $|1\rangle$ and $|2\rangle$ and out of $|0\rangle$. 
The double dot is assumed to be in the strong Coulomb blockade regime 
where the charging energy is larger than the single particle excitation 
energy. Electron tunneling is one--by--one, and the system can effectively
be described by its three states $0$, $1$, $2$, and the empty state $e$ 
before or after one additional electron has tunneled.
The resulting density-matrix equations are
\begin{mathletters}
\begin{eqnarray}
\dot{\rho}_{1,1} &=& \alpha_1\Gamma^0 {\rho}_{0,0}+\Gamma {\rho}_{e,e} 
+2\gamma \rho_{2,2}+ {\rm Im}[\Omega_P (t)\tilde{\rho}_{1,0}] \\
\dot{\rho}_{2,2} &=& \alpha_2\Gamma^0 {\rho}_{0,0}+\Gamma {\rho}_{e,e}
 -2\gamma \rho_{2,2}+ {\rm Im}[\Omega_S(t)\tilde{\rho}_{2,0}] \\
\dot{\rho}_{0,0} &=& -(\Gamma+\Gamma^0) \rho_{0,0} -
{\rm Im}[\Omega_P (t)\tilde{\rho}_{1,0}]
- {\rm Im}[\Omega_S(t)\tilde{\rho}_{2,0}] \\
\dot{\rho}_{e,e} &=& -2\Gamma\rho_{e,e} +\Gamma \rho_{0,0}\\
\dot{\tilde{\rho}}_{1,0} &=& -\left[
\frac{1}{2}\left( \Gamma+\Gamma^0\right) + i\delta_P\right]
\tilde{\rho}_{1,0}
+{i\over 2}\Omega_P(t) (\rho_{0,0}-\rho_{1,1})\nonumber \\
&& -{i\over 2}\Omega_S(t) \tilde{\rho}_{1,2}\\
\dot{\tilde{\rho}}_{2,0} &=& 
-\left[ \frac{1}{2} \left( \Gamma+\Gamma^0\right) +i
\delta_S\right] \tilde{\rho}_{2,0}
+{i\over 2}\Omega_S(t) (\rho_{0,0}-\rho_{2,2})\nonumber\\
&&-{i\over 2}\Omega_P(t) \tilde{\rho}_{1,2}^{*}\\
\dot{\tilde{\rho}}_{1,2} &=& -\left( \gamma +i\delta_R\right)
\tilde{\rho}_{1,2}
+{i\over 2}\Omega_P(t) \tilde{\rho}_{0,2}
-{i\over 2}\Omega_S(t) \tilde{\rho}_{1,0},
\end{eqnarray}
\label{eq:bloch}
\end{mathletters}
where $\alpha_1$ and $\alpha_2=1-\alpha_1$ determine the branching ratio for
the decay of the excited state with rate $\Gamma^0$.

\subsection{Parameters}
Before discussing the numerical results, we 
address typical values of the parameters for lateral quantum dots in GaAs/AlGaAs.
Coulomb charging energies are of a few meV and set the largest energy scale so that
states with more than one additional transport electron can be neglected.
Typical excitation energies  $\delta\varepsilon_i$ from the ground states $|i\rangle$ ($i=1,2$) 
are of the order of a few hundred $\mu$eV which corresponds to microwave frequencies
up to the 100 GHz range. The decay of the excited state $|0\rangle$ then is due to 
acoustic phonon emission at typical rates of $\Gamma^0\approx 10^{9}$ s$^{-1}$.  

Microwave experiments so far have been performed in cw mode only, and we expect the
pulse strengths to be difficult to control. The latter determine, together with the 
two dipole matrix elements for the transitions $|i\rangle\to |0\rangle$, the 
parameters $\Omega_0$ and $\theta$, Eq. (\ref{eq:pulse}).
Fortunately, 
%Typical values for $\Gamma^0$ in GaAs are $10^{9}$ s$^{-1}$ for
%excitation energies of $0.5$ meV.
%Typical microwave frequencies are determined by single particle excitation
%energies $\delta\varepsilon_i$ from the ground states $|i\rangle$ ($i=1,2$) to
%the excited state $|0\rangle$ and depend on the number of electrons in the dots.
%Lock--in techniques and picosecond pulses corresponding to a broad 
%frequency spectrum have already been used to detect the photo--response 
%of dots up to 400 GHz \cite{Blietal98a}. Dots containing around 200 electrons 
%would yield frequencies around $30$ GHz (**** check ****) 
%but can be larger for a smaller number of electrons.  
%Pulse strengths are difficult to control with microwaves but fortunately
control of the strength ratio $\Omega_P/\Omega_S$ is sufficient for the
scheme to work, apart from the requirement to generate pulsed electric fields
${\bf E}_i(t)$ in smooth forms.

Recently, it has been suggested that the dephasing of the optical
coherence for transitions in (electron and hole) quantum dots
is produced by elastic LO-phonon carrier collisions \cite{uskov}.
In our case, such processes result in a much smaller dephasing for ground state
electron superpositions because the quasi-degenerate states forming the
superpositions experience a similar phase shift while scattering. 

Instead, there is strong experi\-mental \cite{FujetalTaretal} and theo\-retical 
\cite{BK99} evidence that the spontaneous emission of 
acoustical phonons is the most dominant process affecting ground state
superpositions $|\psi_f\rangle$ based on the electron charge for sub--Kelvin regime temperatures.
Microscopic coupling constants that determine the dephasing rate $\gamma$ 
of the coherence $\tilde{\rho}_{1,2}$, Eq.(4g), for the different 
dephasing channels (bulk and surface acoustic phonons \cite{DVBK00}) are difficult to calculate,
in particular as there might be additional contributions from non--equilibrium phonons
induced by the microwave radiation itself 
%the dot structure %acting as a transducer for surface acoustic phonons 
\cite{QHEB00preprint}. 
A decisive advantage of coupled dots, however,  is the fact that $\gamma\propto T_c^2$
can be tuned to values which are orders of magnitude smaller than $\Gamma^0$
by varying the tunnel coupling $T_c$ between the dots.

Here, we do not pursue a microscopic approach to the dephasing rate $\gamma$, but
rather solve the dynamics of the dot for given $\gamma$ numerically. As we show now,
this allows us to infer the value of $\gamma$ in a real experiment from another
observable quantity, i.e. the electric current, 
much similar to a recent time--resolved experiment in a superconducting 
system \cite{nakamura}.
Note that due to the Pauli blocking of the leads and the Coulomb blockade,
electrons trapped in the coherent ground state superposition cannot tunnel out
of the dot and no second electron can tunnel in. This is why a 
coupling to external leads of the qubit does not introduce additional 
dephasing channels \cite{BR00}.

%%%%%%%%%%%%%%%%%%%%%%%%%%%%%%%%%%%%%%%%%%%%%%%%%%%%%%%%%%%%%%%%%%
% PLACE PICTURE HERE FOR 4-page version
\begin{figure}[ht]
\setlength{\unitlength}{1in}
%\begin{center}
\begin{picture}(3,3)
%\begin{picture}(4,4)
\put(0,0){\epsfxsize 2.5in \epsfbox{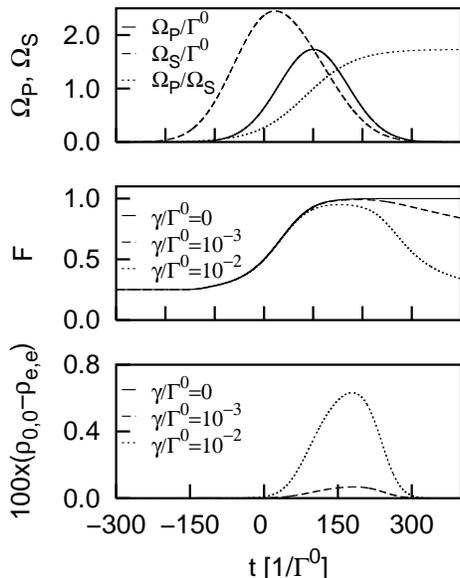}}
%\put(0,0){\epsfxsize 3in \epsfbox{fig1.eps}}
\end{picture}
\caption{
Rabi frequencies, fidelity and electric current as a function of the
interaction time. Parameters of the calculations are: 
$\Omega^o = 2\Gamma^o$, $\Gamma= \Gamma^o/3$, $\theta=\pi/3$, 
$\alpha_1=\alpha_2=1/2$.}
\label{fig_fidelity}
\end{figure}
%%%%%%%%%%%%%%%%%%%%%%%%%%%%%%%%%%%%%%%%%%%%%%%%%%%%%%%%%%%%%%%%%%

\section{Numerical Results}
%ADDED 
We now discuss our results from the numerical solution of the equations of motion, Eq.(\ref{eq:bloch}).
For ideal adiabatic evolution and long-living ground state 
superpositions ($\gamma=0$), the  pulses (\ref{eq:pulse}) 
prepare the dot in the superposition
%\begin{equation}
$
|\psi_f\rangle=\cos\theta |1\rangle - \sin\theta |2\rangle~,
$
%\end{equation}
Eq. (\ref{psif}).  For finite $\gamma$, the adiabatic steering is disturbed 
and the superposition decays into a mixture. 

\subsection{Single Pair of Pulses}
We solved the density-matrix equations
(\ref{eq:bloch}) and calculated the fidelity of the preparation in
 $|\psi_f\rangle$, 
\begin{equation}
F = \langle \psi_f| \rho |\psi_f\rangle
\end{equation}
as a function of the interaction time. Numerical results for $F$ 
for different values of the ground state relaxation rate $\gamma$
are reported in Fig. \ref{fig_fidelity} together with a plot of the 
Rabi frequencies of the pulses, as from Eq. (\ref{eq:pulse}).
In all the numerical calculations $T=\tau=100/\Gamma^o$. 
%ADDED
This corresponds to
a pulse of the order of one tenth microsecond, resulting in a frequency 
dispersion of the order of 10 MHz. Ground state splittings larger than 
that are easily achieved in quantum dots by an appropriate tuning via 
gate--voltages, so that the assumption that each microwave field couples only 
one transition is fully justified.
Furthermore the ground state splitting has to be larger than the excited
state width, to avoid unwanted couplings.

In the absence of ground state relaxation processes, the evolution is 
analogous to what is known from atomic physics: the interaction
with the light leads to the preparation of the dot in the desired 
superposition of states. After the pulse sequence the dot stays
in this superposition. For small but nonzero relaxation rate ($\gamma=0.001\Gamma^o$), 
one recognizes that the microwave driving still results in the 
preparation of the dot in $|\psi_f\rangle$, but after the pulse sequence
the fidelity degrades rapidly. For a larger relaxation rate 
($\gamma\simeq 0.01\Gamma^o$), also the driving into the superposition is 
disturbed, and the preparation in $|\psi_f\rangle$ is never complete.
Furthermore, there is a fast degradation of the fidelity after the pulse
sequence.

We now show how to get access to the ground state evolution and to $\gamma$
by a current measurement. 
As shown in Fig. \ref{fig_fidelity}, it is in principle 
possible to get insight on the ground state dynamics by monitoring 
the time--dependent electric current through the dot
\begin{equation}
I(t)=-e\Gamma  [\rho_{0,0}(t)-\rho_{e,e}(t)],
\end{equation}
due to the flow of electrons with charge $-e<0$ through
the tunnel barrier connecting the dot into the right reservoir.
However, for small relaxation rates the current through the dot is weak;
%*** CHANGED and therefore difficult to monitor. 
note that the scale in Fig.\ref{fig_fidelity}
is blown up by a factor 100 and that a current $-e\Gamma/100$ with 
$\Gamma=10^{-9}$ s$^{-1}$ corresponds to 1.6 pA.

\subsection{Double Pulse Sequence}
A more sensitive detection is 
%*** CHANGED  therefore 
obtained by letting the system evolve 
freely, and then applying the microwave radiation once again. The two pulses 
that have been used for the preparation of the state $|\psi_f\rangle$ are
then applied simultaneously  
%%%%CHANGED
at a second time $\Delta t>0$ (i.e. after the first pair  of pulses) 
%%%%%%%%%
with 
\begin{mathletters}
\begin{eqnarray}
\Omega_P(t)&=&\Omega_p \sin \theta e^{-(t-\Delta t)^2/T_p^2},\\
\Omega_S(t)&=&\Omega_p \cos (\theta + \phi) e^{-(t-\Delta t)^2/T_p^2}
\end{eqnarray}
\label{eq:probe}
\end{mathletters}
and the ratio of their amplitudes 
corresponding to $|\psi_f\rangle$ ($\phi=0$) or to its orthogonal state
($\phi=\pi$).

%%%%%%%%%%%%%%%%%%%%%%%%%%%%%%%%%%%%%%%%%%%%%%%%%%%%%%%%%%%%%%%%%%
\begin{figure}[ht]
\setlength{\unitlength}{1in}
%\begin{picture}(4,3.5)
\begin{picture}(4,3.5)
\put(0,0){\epsfxsize 2.5in \epsfbox{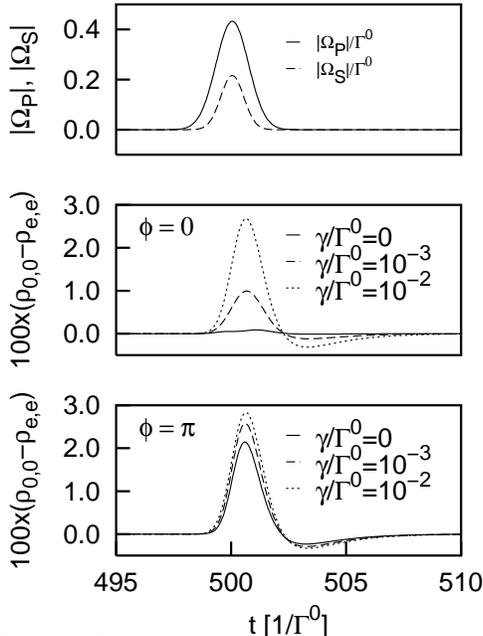}}
\end{picture}
\caption{
Rabi frequencies and electric current as a function of the   
interaction time. Parameters of the calculations are:
$\Omega_p = 0.5\Gamma^o$, $T_p=1/\Gamma^o$ and $\Delta t = 500/\Gamma^o$.}
\label{fig_probe}
\end{figure}
\begin{figure}[ht]
\begin{center}
\mbox{\epsfxsize 3in \epsfbox{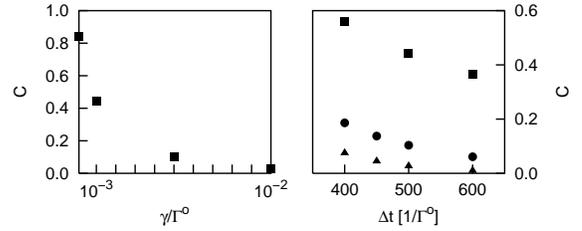}}
\end{center}
\caption{
Contrast $C$ as a function of the ground state relaxation rate (left) and
time delay $\Delta t$ (right) of the probe pulses.
In the left plot squares correspond to $\Delta t= 500/\Gamma^0$. 
In the right plot squares correspond to $\gamma/\Gamma^0=10^{-3}$, 
circles to $5\cdot 10^{-3}$ and triangles to $10^{-2}$.
Parameters for the probe pulses are the same as for Fig. \ref{fig_probe} }
\label{contrast}
\end{figure}

%%%%%%%%%%%%%%%%%%%%%%%%%%%%%%%%%%%%%%%%%%%%%%%%%%%%%%%%%%%%%%%%%%
As shown in Fig. \ref{fig_probe}, if $\gamma=0$ and $\phi=0$, nothing 
happens: the dot stays in the state $|\psi_f\rangle$ and the subsequent
application of the probe pulses does not produce any current through the
dot. On the contrary, for $\phi=\pi$ the probe pulses are in anti--phase with 
the ground state superposition and a large current follows. 
For a nonzero relaxation rate $\gamma$ the superposition decays into a 
mixture on a time scale $1/\gamma$ and therefore the application 
of the probe pulses results in a current through the dot both for $\phi=0$
and $\phi=\pi$. 
The larger the relaxation rate $\gamma$, the less sensitive
is the current on the relative phase $\phi$ of the probe pulses. Therefore,
the contrast $C$
\begin{equation}
C=\frac{I_{max}(\phi=\pi)-I_{max}(\phi=0)}{I_{max}(\phi=\pi)+I_{max}(\phi=0)}
\end{equation}
is a good measure of the ground state relaxation rate $\gamma$, as shown in
Fig. \ref{contrast}.

% CHANGED
%%%%%%%%%%%%%%%%%%%%%%%%%%%%%%%%%%%%%%%%%%%%%%%%%%%%%%%%%%%%%%%%%%

The relaxation rate $\gamma$ thus can be extracted experimentally in the following way:
the dot is prepared in $|\psi_f\rangle$ as described and then
probed at the instant $\Delta t$ by pulses with a relative phase $\phi=0$.
The cycle preparation/probing is then repeated for $\phi=\pi$.
Different measurement of $C$ in this way are made for various choices of
$\Delta t$. A plot of the contrast $C$ as a function of $\Delta t$ 
then allows the determination of $\gamma$.

\section{Conclusions}
In conclusion, we have shown that the time--evolution of 
double dots in the regime of strong Coulomb blockade under coherent microwave
pulses provides a scheme that allows both preparation and analysis of eigenstate superpositions 
$|\psi_f\rangle$.
A `read--out' current pulse provides the information about the 
superposition dephasing rate $\gamma$. In double dots it is possible to fine--tune $\gamma$ by varying 
the interdot--coupling by gate--voltages, and to prepare  $|\psi_f\rangle$ as a dark state
that is protected deeply below the Fermi seas of the contact reservoirs by the Pauli principle and the 
Coulomb blockade effect.
This suggests that it might be worthwile to further investigate 
(among other proposals based on, e.g., the electron spin \cite{LV98})
such charge--based qubits that have been proven
to be accessible to transport spectroscopy.

\bibliographystyle{prsty}

\end{document}